\documentclass[11pt,a4paper,twoside]{article}
\usepackage[T1]{fontenc}
\usepackage[latin1]{inputenc}
\usepackage{amsmath,amssymb,amsthm}
\usepackage{graphicx,color}
\usepackage{parskip}
\def\be{\begin{equation}}
\def\ee{\end{equation}}
\def\bea{\begin{eqnarray}}
\def\eea{\end{eqnarray}}
\newcommand{\rmd}{\mbox{d}}

 \newcommand{\ket}[1]{|\kern.3ex#1\kern.3ex\rangle}
 \newcommand{\bra}[1]{\langle\kern.3ex #1 \kern.3ex|}
 \newcommand{\mean}[1]{\langle#1 \rangle} 

\begin{document}
\title{Random Aharonov-Bohm vortices and some funny families of integrals}
\author{St\'ephane  OUVRY\footnote{ouvry@lptms.u-psud.fr}$^a$}
\date{February 11 2005}



\maketitle
{\small

\noindent\hspace{1cm}
$^a$
\begin{minipage}[t]{12.5cm}
 LPTMS, UMR 8626-CNRS\\ B\^at 100, Universit\'e d'Orsay, F-91405 Orsay, France
\end{minipage}
}

\abstract {A review of the random magnetic impurity model, introduced in the
context of the Quantum Hall effect, is presented. It models an electron moving
in a plane and coupled to random Aharonov-Bohm vortices carrying a fraction
of the flux quantum. Recent results on its perturbative expansion are
given. In particular, some funny families of integrals show up to be related to
the Riemann $\zeta(3)$ and $\zeta(2)$.}


\section{Introduction :}

The single Aharonov Bohm (A-B) flux line (infinitely thin and impenetrable
vortex) piercing the plane at a given point was considered in a seminal paper
in 1959 \cite{AB}. A more complex system consists of several A-B lines
piercing the plane at different points \cite{2AB}. A simplification arises
when the locations of the punctures are random : consider the Random Magnetic
Impurity model \cite{nous} introduced in the context of the Integer Quantum
Hall Effect. It consists of an electron coupled to a Poissonian distribution
of A-B lines (the so-called magnetic impurities) having a mean density $\rho$
and carrying a fraction $\alpha$ of the electron flux quantum $\Phi_o$.
Periodicity ($\alpha\in[0,1]$) and symmetry considerations (no priviledged
orientation to the plane) allow to take $\alpha\in[0,1/2]$. The Poissonian
disorder defines the infinitesimal probability $dP(N,\vec r\,'_1, \vec
r\,'_2,...,\vec r\,'_N)$ of finding $N$ impurities at position $\vec r\,'_1,\vec
r\,'_2,...,\vec r\,'_N$ 
\be\label{poisson} dP(N,\vec r\,'_1, \vec r\,'_2,...,\vec r\,'_N) =e^{-\rho V}{(\rho
V)^N\over N!}\prod_{i=1}^N{d\vec r\,'_i\over V}\nonumber\ee
where $\rho$ is the mean impurity density : in the thermodynamic limit
$N,V\to\infty$ ($V$ the area of the plane), $\rho=<N>/V$.
 
The model has two parameters, $\alpha$ and $\rho$. 
Periodicity $\alpha\to \alpha+1$
and symmetry around $\alpha=1/2$ in the spectrum imply that the density of states and the partition
function are invariant under $\alpha\to 1-\alpha$, thus depend in fact on
$\rho$ and $\alpha(1-\alpha)$.

It has been shown via path integral random walk
simulations that, when $\alpha\to 0$, the average density of states of the
electron narrows down to the Landau density of states for the average magnetic
field $<B>=\rho \alpha \Phi_o$ with broadened Landau levels (weak disorder).
On the contrary, when $\alpha\to 1/2$, the density of states has no Landau
level oscillations and rather exhibits a Lifschitz tail at the bottom of the
spectrum (strong disorder).

In the path integral formulation, 
the average partition function  rewrites as an average over $ C$, the set of
closed Brownian curves of a given length $t$ (the inverse temperature) 
\[ <Z>=Z_o<e^{\rho\sum_nS_n(e^{i 2\pi \alpha
n}-1)}>_{\{C\}}\] 
where $S_n$ is the arithmetic area of the $n$-winding sector
(i.e. the sector wounded around  $n$ times by the path) of a given path in
$\{C\}$ ($Z_o$ is the free partition function). It amounts to say that
the Poissonian A-B lines couple to the $S_n$'s, a different (intermediate)
situation from the single A-B line, which couples to the angle spanned by the
path around it, and from the homogeneous magnetic field, which couples to the
algebraic area enclosed by the path. The average partition function rewrites
as
\[
<Z>=Z_o\int e^{-\rho t(S+iA)}P(S,A)dSdA\] where ${{S}}={2\over
t}\sum_nS_n\sin^2(\pi\alpha n)$ and $ {A}={1\over t}\sum_nS_n \sin(2\pi\alpha
n)$ are random Brownian loop variables and $P(S,A)$ is their joint 
probability distribution. Since \cite{werner}  $S_n$ scales like $t$ 
-in fact
$<S_n>=t/(2\pi n^2)$, even more, for $n$ sufficiently large, $n^2S_n\to
<n^2S_n>=t/(2\pi)$-, the variables $S$ and $A$ are indeed $t$ independent.

 In fact, very little is known  on the joint distribution
function $P(S,A)$:

i) when $\alpha\to 0,  <Z>\to Z_o<e^{i<B>\sum_n nS_n}>_{\{C\}}$,  
which is,as expected,  the partition
function  for the homogeneous mean magnetic field $<B>$,
since  $\sum nS_n$ is indeed the
algebraic area enclosed by the path. 

ii) when $\alpha\to 1/2, <Z>\to Z_o<e^{-2\rho  \sum_{n\,{\rm odd}} S_n}>_{\{C\}}$,
implying that, for the average density of states (in terms of $\rho_o(E)$
the free density of states)
\[<\rho(E)>=\rho_o(E)\int_0^{{E\over
\rho}}P(S')dS'\] 
Here $P(S')$ is the probability distribution for the random
variable $S'={2\over t}\sum_n S_n$, $n$ odd.

Turning now to a quantum mechanical formulation, the average partition
function is

\bea <Z>=e^{-\rho V}\sum_N{(\rho V)^N\over N!}<Z_N>\quad\quad\quad\quad\nonumber\\=Z_o\bigg(1+\rho V({<Z_1>\over Z_o}-1)+{1\over 2!}(\rho V)^2(
{<Z_2>\over Z_o}-2{<Z_1>\over Z_o}+1)\nonumber\\+{1\over 3!}(\rho V)^3({<Z_3>\over Z_o}-
3{<Z_2>\over Z_o}
+3{<Z_1>\over Z_o}-1)+...\bigg)\nonumber\eea
where $<Z_N>$ is  the average $N$ impurity partition function  
\be <Z_N>=\int\prod_{i=1}^N{d\vec r\,'_i\over V}Tr e^{-\beta H_N}\nonumber\ee
for the  $N$-impurity quantum Hamiltonian  $H_N$.

The partition function of the mean magnetic field $\mean{B}=\rho\alpha\phi_o$
is reproduced as a power series in $(\rho\alpha)^n$ (mean field approximation
where the local magnetic field $B(\vec r)=\phi\sum_{i=1}^N \delta(\vec r-\vec
r\,'_i)$ is replaced by its mean value). Perturbative corrections
$\rho^n\alpha^m, m>n$ to the mean magnetic field expansion originate from
disorder: in the 1-impurity case the correction is known \cite{georg}
$<Z_1>=Z_o+\alpha(\alpha-1)/2$; in the 2-impurity case one encounters non
trivial Feynman diagrams at order $\rho^2\alpha^4$, i.e. an electron
interacting with 2 impurities 4 times, at order $\rho^2\alpha^6$, i.e. an
electron interacting with 2 impurities 6 times, etc...

In the symmetric gauge and in
configuration space
\be\label{11} H_N={1\over 2} \left(\vec p - \sum_{i=1}^N\alpha{\vec
k\times(\vec r-\vec r\,'_i) \over (\vec r -\vec r\,'_i)^2} \right)^2
+\pi\alpha\sum_{i=1}^N \delta(\vec r-\vec r\,'_i)\nonumber \ee
The nonunitary wavefunction redefinition 
\be\label{new1} \psi(\vec r)=\prod_{i=1}^N\vert\vec r-\vec
r\,'_i\vert^{\alpha} \tilde{\psi}(\vec r)\nonumber \ee 
results in the Hamiltonian $\tilde{H}_N$ acting on $\tilde{\psi}$
($z$ complex coordinates) 
\be \label{101}
\tilde{H}_N=-2\partial_{\bar z}\partial_z -2\alpha\sum_{i=1}^N {1\over \bar
z-\bar z'_i}\partial_z =H_o+\tilde V_N(z)\nonumber\ee 
where quadratic interactions in the
vector potential have disappeared.

The perturbative expansion for the thermal propagator $\tilde{G}_\beta=
e^{-\beta\tilde{H}_N}$
at order $n$ is
\bea\label{pertexp}
\tilde{G}^{(n)}_\beta =
(-1)^n \int_0^\beta d\beta_1 \int_0^{\beta_1} d\beta_2\cdots 
\int_0^{\beta_{n-1}} d\beta_n \nonumber\\
G^{(o)}_{\beta-\beta_1}
{\tilde V_N} G^{(o)}_{\beta_1-\beta_2} 
\cdots
{\tilde V_N}G^{(o)}_{\beta_n}\nonumber
\eea
where $G^{(o)}_{\beta}=e^{-\beta{H_o}}$ is the free 
 propagator.
The partition function at order $n$ follows as

\bea Z_N^{(n)}=Tr {G}^{(n)}_\beta=Tr \tilde{G}^{(n)}_\beta
=(-1)^n Tr \int_0^\beta d\beta_1 \int_0^{\beta_1} d\beta_2\cdots 
\int_0^{\beta_{n-1}} d\beta_n \nonumber\\
{\tilde V_N} G^{(o)}_{\beta_1-\beta_2} 
\cdots
{\tilde V_N} G^{(o)}_{\beta+\beta_n-\beta_1}\nonumber
\eea
   
In  the configuration space one has

\bea Z_N^{(n)}=(-1)^n \int_0^\beta d\beta_1 \int_0^{\beta_1} d\beta_2\cdots 
\int_0^{\beta_{n-1}} d\beta_n \int d^2z_1\cdots \int d^2 z_n \nonumber\\
{\tilde V}_N(z_1) G^{(o)}_{\beta_1-\beta_2}(z_1,z_2) 
\cdots
{\tilde V}_N(z_n) G^{(o)}_{\beta+\beta_n-\beta_1}(z_n,z_1)
\nonumber
\eea

\[G_{\beta}^{(o)}(z_1,z_2)=\bra{z_1}
e^{-\beta H_o}\ket{z_2}={1\over 2\pi\beta}e^{ -{|z_1-z_2|^2\over 2\beta}}\]

\[\tilde V_N(z_1)=\bra{z_1}
\tilde V_N\ket{z_1}= -2\alpha\sum_{i=1}^N {1\over \bar
z_1-\bar z'_i}\partial_{z_1}\]
with the average over disorder
\begin{eqnarray}\label{meanrules} 
\int d^2z'_i {1\over \bar z_1-\bar z'_i}&=&\pi z_1\\
\int d^2z'_i{1\over \bar z_1-\bar z'_i}
{1\over \bar z_2-\bar z'_i}&=&\pi ({z_1\over \bar z_2-\bar z_1}
+{z_2\over \bar z_1-\bar z_2})\quad\\
&\vdots&\nonumber
\end{eqnarray}
Note that if at first order in $\alpha$ an ambiguity arises since ${\tilde V}(z_1)$ is a differential
operator acting on the constant $G^{(o)}_{\beta}(z_1,z_1)$, this ambiguity can be lifted  by a long distance harmonic
well regularization or by  other means (see for example \cite{nous}).

In momentum space one has ($p$ complex coordinate)
  
\[G_{\beta}^{(o)}(p_1,p_2)=\bra{p_1}
e^{-\beta H_o }\ket{p_2}=e^{-\beta{|p_1|^2\over 2}}{\delta^2(p_1- p_2)}\]

\[\bra{p_1}
\tilde V_N\ket{p_2}= -2\alpha\sum_{i=1}^N {\bar p_2\over \bar p_1-\bar p_2}
e^{{i\over 2}(p_1-p_2)\bar z'_i+h.c.}\]
with the average over disorder implying momentum conservation.

With these perturbative Feynman rules at hand, one can proceed with the
perturbative expansion of the average partition function. The first non
trivial diagram occurs at order $\rho^2\alpha^4$. Denoting the temperature
differences as $a=\beta_1-\beta_2, b=\beta_2-\beta_3, c=\beta_3-\beta_4, d=
\beta+\beta_4-\beta_1$, the momentum integration yields an integral over the
temperatures
\[{\rm diag}(\rho^2\alpha^4)= {1\over \beta^2}\int_0^{\beta}d\beta_1\int_0^{\beta_1}d\beta_2
\int_0^{\beta_2}d\beta_3\int_0^{\beta_3}d\beta_4\bigg(
{2\over \beta}-{(a+c)(b+d)\over abc +bcd+cda+dab}\bigg)\] 
Integrating on
$\beta_4$, $\beta_3$, etc...,  leads for the non trivial part to \cite{nous}
\be\label{ninn} \int_0^{\beta}d\beta_1\int_0^{\beta_1}d\beta_2
\int_0^{\beta_2}d\beta_3\int_0^{\beta_3}d\beta_4{(a+c)(b+d)\over abc
+bcd+cda+dab}=\beta^3{1+\tilde\zeta(3)\over 16}\ee
where  $\tilde\zeta(3)=7\zeta(3)/2$.
One can go a bit further : first rewrite this integral as 
\[ \int_{0}^{\beta}d a\,\int_0^{\beta-a} d b\,\int_0^{\beta-a-b} d c\, 
{d(a+c)(b+d)\over abc+bcd+cda+dab}\]
where $d=\beta-(a+b+c)$ is understood. This is
\[ \int_{a,b,c,d=0}^{\infty}d a\, d b\, d c\, d d\, {d(a+c)(b+d)\over abc+bcd+cda+dab}
\delta(\beta-(a+b+c+d))\]
or, using obvious symmetries,
\[ \beta\int_{a,b,c,d=0}^{\infty}d a\, d b\, d c\, d d\, {ab\over abc+bcd+cda+dab}
\delta(\beta-(a+b+c+d))\]
Therefore (\ref{ninn})becomes
\be\label{nonn} \int_{a,b,c,d=0}^{\infty}d a\, d b\, d c\, d d\, {ab\over abc+bcd+cda+dab}
\delta(\beta-(a+b+c+d))=\beta^2{1+\tilde\zeta(3)\over 16}\ee
Then Laplace transform (\ref{nonn})  ($ \int_0^{\infty}d \beta
e^{-\beta s}...$) to obtain
\[ 
{1\over s^3}\int_{a,b,c,d=0}^{\infty}d a\, d b\, d c\, d d\, {1\over cd({1\over a}+{1\over b}+{1\over c}+{1\over d}
)} e^{-s(a+b+c+d)}={2\over s^3}{1+\tilde\zeta(3)\over 16}\]
Fix $s=1$ and, in the denominator, exponentiate ${1\over a}+{1\over b}+{1\over c}+{1\over d}$: 
\[ 
 \int_{a,b,c,d=0}^{\infty}d a\, d b\, d c\, d d\, \int_0^{\infty}
d t\, {1\over cd}
e^{-(a+b+c+d)-t({1\over a}+{1\over b}+{1\over c}+{1\over d})}=2{1+\tilde\zeta(3)\over 16}\]
Change  variable $u=2\sqrt{t}$ and use 
 \be\label{} \int_0^{\infty}a^{n-1}e^{ -a-{t\over a}}
=2K_{n}(u)({u\over 2})^{n}\nonumber
\ee
where  the $K_{\nu}(u)$'s are modified Bessel functions \cite{grad},
to  finally obtain
\be\label{conj}\int_{0}^{\infty}{ud u\,} (uK_1(u))^2K_0(u)^2=
{1+{\tilde\zeta(3)} \over 16}\ee
which also been derived in a different context \cite{tcheque}.

It  can be generalized \cite{nous'} :  the family of integrals 

\[p_n(0000)=\int_{0}^{\infty}\rmd u\,  u^{n+1}
K_{0}(u)^4\quad n\ge 0\]
\[ p_n(0011)=\int_{0}^{\infty}\rmd u\,  u^{n+1}
K_{0}(u)^2K_{1}(u)^2\quad n\ge 2\]
\[p_n(1111)=\int_{0}^{\infty}\rmd u\,  u^{n+1}
K_{1}(u)^4\quad n\ge 4\]
($n$ even)

\noindent and 
\[ i_n(0001)=\int_{0}^{\infty}\rmd u\,  u^{n+1}
K_{0}(u)^3K_{1}(u)\quad n\ge 1\]
\[ i_n(0111)=\int_{0}^{\infty}\rmd u\,  u^{n+1}
K_{0}(u)K_{1}(u)^3\quad n\ge 3\]
($n$ odd)
\noindent can be shown to be  again related to $\zeta(3)$, i.e. 
of the form (\ref{conj}).
To show this, integrate by parts 
(use $\rmd K_{0}(u)/\rmd u\,=-K_{1}(u)$ and $\rmd
(uK_{1}(u))/\rmd u\,=-uK_{0}(u)$)
to obtain
\[ p_n(0000)={4\over n+2}i_{n+1}(0001) \quad n\ge 0\]
\[
p_n(0011)={2\over n}(i_{n+1}(0001)+i_{n+1}(0111))\quad n\ge 2\]
\[ p_n(1111)={4\over n-2}i_{n+1}(0111)\quad n\ge 4\]
which implies
\[ 2np_n(0011)=(n+2)p_n(0000)+(n-2)p_n(1111)\quad n\ge 4\]
Therefore $i_{n+1}(0001), i_{n+1}(0111), {\rm and} p_n(0011)$ are obtained 
from $p_n(0000)$ and $p_n(1111)$ which remain to be determined:  
further integration by parts gives
\[ i_n(0111)={1\over n-1}((p_{n+1}(1111)+3p_{n+1}(0011))\quad n\ge 3\]
\[ i_n(0001)={1\over n+1}((p_{n+1}(0000)+3p_{n+1}(0011))\quad n\ge 1\]
thus the recurrence relation acting on a 2 dimensionnal vector space
\be\label{rec}
\left(
\begin{array}{c}
p_{n+2}(0000)\\ 
p_{n+2}(1111)\end{array}\right)
={1\over 2^5(n+2)}
\left(
\begin{array}{cc}
(n+2)^2(5n+4) & -3n^2(n-2)\\  
 -3(n+2)^2(n+4) & n(n-2)(5n+16) 
\end{array}\right)
\left(
\begin{array}{c}
p_{n}(0000)\\ 
p_{n}(1111)\end{array}\right)
\ee
for $ n\ge 4$.
For $0\le n<4$ direct computations (one has to rewrite these simple integrals in terms of multiple integrals) give
\noindent
\begin{eqnarray}p_0(0000)&=& {{\tilde\zeta(3)}\over 2^2}\nonumber\\
 i_1(0001)&=& {{\tilde\zeta(3)}\over 2^3 }\nonumber\\
  p_2(0000)&=&{-3+{\tilde\zeta(3)}\over 2^4}\nonumber  \\
 p_2(0011)&=&{1+{\tilde\zeta(3)}\over 2^4 }\quad {\rm known} \equiv (\ref{conj}) \nonumber\\
 i_3(0001)&=&{-3+{\tilde\zeta(3)}\over 2^4 }\nonumber\\
 i_3(0111)&=&{1\over 2^2}\quad {\rm obvious}\nonumber
\end{eqnarray} and the $n=4$ initial conditions for the recurrence are
\begin{eqnarray}\label{init}
p_{4}(0000)&=&
  {-3^3+7{\tilde\zeta(3)}\over 2^6}\nonumber\\
 p_4(1111)&=&{53-3^2{\tilde\zeta(3)}\over 2^6}
\end{eqnarray}
For example, in terms of the corresponding multiple integral, 
$p_0(0000)=\tilde\zeta(3)/2^2$ is nothing but
\be\label{}\int_{a,b,c,d=0}^{\infty}d a\, d b\, d c\, d d\, {1\over
abc+bcd+cda+dab} e^{-(a+b+c+d)}=7\zeta(3)\nonumber\ee
 If one sets $n=2k$ and defines
$q_k(0)=p_{2k}(0000)/(2k)!$ and
 $q_k(1)=p_{2k}(1111)/(2k)!$
 the recurrence (\ref{rec}) becomes 
\be\label{recbis}
\left(
\begin{array}{c}
q_{k+1}(0)\\ 
q_{k+1}(1)\end{array}\right)
={1\over 2^4(2k+1)}
\left(
\begin{array}{cc}
(5k+2) & {-3k^2(k-1)/ (k+1)^2}\\  
 -3(k+2) & {k(k-1)(5k+8)/ (k+1)^2} 
\end{array}\right)
\left(
\begin{array}{c}
q_{k}(0)\\ 
q_{k}(1)\end{array}\right)
\ee
$k\ge 2$. 
In the asymptotic regime $k\to\infty$  
\bea
{1\over 2^5}
\left(
\begin{array}{cc}
5 & -3\\  
-3 & 5 
\end{array}\right)
\nonumber\eea
 diagonalizes as 
\[
{1\over \sqrt{2}}
\left(
\begin{array}{cc}
1 & 1\\
-1 & 1 
\end{array}\right)
{1\over 2^5}\left(
\begin{array}{cc}
5 & -3\\  
 -3 & 5 
\end{array}\right)
{1\over \sqrt{2}}\left(
\begin{array}{cc}
1&-1\\
1&1 
\end{array}\right)
={1\over 2^4}
\left(
\begin{array}{cc}
1 & 0\\  
 0 & 4 
\end{array}\right)
\nonumber\] with eigenvalues $1/4$ and $1/16$. It is a remarkable fact that
the initial conditions (\ref{init}), that is for the $q$'s 
\[q_2(0)={-27+7\tilde\zeta(3)\over(3)(2^8)},\quad 
q_2(1)={53-9\tilde\zeta(3)\over(3)(2^8)}\] are   such that
the asymptotic behavior of $ q_k(0)$ and $q_k(1)$ happens to be governed by the
smallest eigenvalue, i.e. $\lim_{k\to\infty} q_k(0)\simeq
(1/16)^{k-2}$ and $\lim_{k\to\infty} q_k(1)\simeq (1/16)^{k-2}$.

This is not the end of the story:  pushing the perturbative expansion at order
$\rho^2\alpha^6$ \cite{inprep} one encounters the integral
$\int_{0}^{\infty}\rmd u\, I_1(u)K_{0}(u)^2(uK_1(u))$, where $I_1(u)$ is again
a modified Bessel function \cite{grad}. This integral is found to be
\be\label{nann} \int_{0}^{\infty}\rmd u\,  
I_1(u)K_{0}(u)^2(uK_1(u))={\zeta(2)\over 8}\ee
It is not an isolated case:
the family of integrals
\[{} p_n^0(000)=\int_{0}^{\infty}\rmd u\,  u^{n}
(I_0(u)u)K_{0}(u)^3u^2\]
\[{} p_n^1(000)=\int_{0}^{\infty}\rmd u\,  u^{n}
I_1(u)K_{0}(u)^3u^2\]
\[{} p_n^0(001)=\int_{0}^{\infty}\rmd u\,  u^{n}
(I_0(u)u)K_{0}(u)^2(uK_1(u))\]
\[{} p_n^1(001)=\int_{0}^{\infty}\rmd u\,  u^{n}
I_1(u)K_{0}(u)^2(uK_1(u))\]
\[{} p_n^0(011)=\int_{0}^{\infty}\rmd u\,  u^{n}
(I_0(u)u)K_{0}(u)(uK_1(u))^2\]
\[{} p_n^1(011)=\int_{0}^{\infty}\rmd u\,  u^{n}
I_1(u)K_{0}(u)(uK_1(u))^2\]
\[{} p_n^0(111)=\int_{0}^{\infty}\rmd u\,  u^{n}
(I_0(u)u)(uK_1(u))^3\]
\[{} p_n^1(111)=\int_{0}^{\infty}\rmd u\,  u^{n}
I_1(u)(uK_1(u))^3\]
($n\ge 0, n$ even)
can be shown to be  also related\footnote{One also has 
\[ \int_{0}^{\infty}\rmd u\,  
(I_0(u)u)K_{0}(u)^3=3\zeta(2)/8 \] but, surprisingly,
$\int_{0}^{\infty}\rmd u\,
I_1(u)K_{0}(u)^3$ is not related to $\zeta(2)$.} 
to $\zeta(2)$ (note that (\ref{nann}) is
$p_0^1(001)=\zeta(2)/8$).

Indeed, integration by part (this exercice is left 
to the interested reader, use ${dI_0(u)\over du}
=I_1(u)$ and ${d(uI_1(u))\over du}
=uI_0(u)$)
 gives \cite{guil} that
\be\label{recter}
\left(
\begin{array}{c}
p^0_{n+2}(000)\\ 
p^1_{n+2}(111)\end{array}\right)
={1\over 2^5(n+4)}
\left(
\begin{array}{cc}
(n+4)^2(5n+14) & 3(n+2)^2(n)\\  
 3(n+4)^2(n+6) & (n+2)(n)(5n+26) 
\end{array}\right)
\left(
\begin{array}{c}
p^0_{n}(000)\\ 
p^1_{n}(111)\end{array}\right)
\ee
and that all other integrals are given in terms of $p^0_{n}(000)$ and $p^1_{n}(111)$.
For example
\bea
2^5(n+3)(n+4)p^0_{n+2}(011)=&+&8(n+2)^2(n+4)^2p^0_{n}(011)\nonumber \\&-&(n+4)^2(3n^2+15n+14)p^0_n(000)\nonumber
\\ &+&n(n+2)(3n^2+27n+58)p^1_n(111)\nonumber\eea
Amusingly enough, the recurrence (\ref{recter}) is identical to (\ref{rec}) up to $n\to n+2$
and  to the off-diagonal global sign, the origin of which being
easily tractable to the sign difference in the differentiation relations 
between $I_0$ and $I_1$ on the one hand, and
$K_0$ and $K_1$ on the other hand.

Setting as above $n=2k, k\ge 0$ and redifining
\be\label{} q^0_k(0)={p^0_{2k}(000)\over (2(k+1))!}\nonumber\ee 
\be\label{} q^1_k(1)={p^1_{2k}(111)\over (2(k+1))!}\nonumber\ee
the recurrence  becomes 
\be\label{}
\left(
\begin{array}{c}
q^0_{k+1}(0)\\ 
q^1_{k+1}(1)\end{array}\right)
={1\over 2^4(2k+3)}
\left(
\begin{array}{cc}
(5k+7) & {3k(k+1)^2/ (k+2)^2}\\  
 3(k+3) & {k(k+1)(5k+13)/ (k+2)^2} 
\end{array}\right)
\left(
\begin{array}{c}
q^0_{k}(0)\\ q^1_{k}(1)\end{array}\right)\nonumber \ee 
with again  the eigenvalues $1/4$ and $1/(16)$  for the asymptotic matrix.
The initial
conditions are (by direct computation)
\be q_0^0(0)= {\tilde\zeta(2)\nonumber}\ee 
\be q_0^1(0)=-{1\over 8}+3\tilde\zeta(2)\nonumber\ee
where $\tilde\zeta(2)=3\zeta(2)/2^6$.

To conclude a few remarks can be made: 

\noindent -in both recurrences, the building blocks, $\tilde\zeta(2)=3\zeta(2)/2^6$ and
$\tilde\zeta(3)=7\zeta(3)/2$, are not pure $\zeta(n)$'s, but 
rather
$(2^n-1)\zeta(n)$'s. This indicates that one deals with alternated Euler sums,
\be\zeta_a(n)=\sum_{p=1}^{\infty} {(-1)^p\over p^n}\nonumber\ee
or 
\be \tilde\zeta(n)={\zeta(n)+\zeta_a(n)\over 2}={2^n-1\over 2^n}\zeta(n)\nonumber\ee
-it has been impossible so far to generalize to $\zeta(4),\zeta(5),..$ in
terms of simple integrals on the product of modified Bessel
functions\footnote{Note that integrals on a product of four modified Bessel  functions seem so far needed:  for example
\[ \int_0^\infty {\rm d}u\, u K_0(u)^3 
={3\over 2}\sum_{p=0}^\infty \frac{1}{(3p+1)^2}
-\frac{2}{3}\zeta(2)\]
is not simply given in terms of $\zeta$'s.}. May be one should look at double
nested integrals as indicated \cite{inprep} by computation at order
$\rho^2\alpha^6$.


\begin{thebibliography}{99}

\bibitem{AB} Y. Aharonov and D. Bohm, Phys. Rev. 115 (1959) 485; for an earlier
description of the same effect see: W. Ehrenberg, R. W. Siday, Proc. Phys.
Soc. London, sec. B 62 (1949) 8

\bibitem{2AB}  P. Stovicek, Phys. Lett. A 142 (1989) 5; S. Mashkevich, J. Myrheim and S. Ouvry, Phys. Lett. A 330 (2004)
41

\bibitem{nous} J. Desbois, C. Furtlehner and S. Ouvry, Nuclear Physics B [FS]
{453} (1995) 759

\bibitem{werner} Alain Comtet, Jean ~Desbois and St\'ephane ~Ouvry, J. Phys. A
23 (1990) 3563; Wendelin Werner, Th\`ese, Universit\'e Paris 7 (1993) and
Probability Theory and Related Fields (1994) 111

\bibitem{georg} Alain Comtet, Yvon Georgelin and St\'ephane ~Ouvry, J. Phys.
A: Math. Gen. 22 (1989) 3917


\bibitem{grad} see Gradshteyn and Ryzhik, Table of Integrals, Series and
Products, Fifth Edition, Academic Press, 8.432 1. and 8.432 6

\bibitem{tcheque} L. Samaj and I. Travenec, J. Stat. Phys 101 (2000) 713

\bibitem{nous'} Cyril Furtlehner and St\'ephane Ouvry, math-ph/0306004,
``Integrals involving four Macdonald functions and their relation to
$7\zeta(3)/2$''

\bibitem{inprep} Cyril Furtlehner and St\'ephane Ouvry, in preparation

\bibitem{guil} This was shown in collaboration with Guillaume de Montlaur, stage de Ma\^\i trise Paris VI (summer 2004)


\end{thebibliography}
\end{document}